\documentclass[showpacs,preprintnumbers,superscriptaddress,preprint]{revtex4-1}
\usepackage{amsmath,amssymb}% 
\usepackage{graphicx}% Include figure files
\usepackage{dcolumn}% Align table columns on decimal point
\usepackage{bm}% bold math

\begin{document}
%Title of paper
\title{Magnetization scaling in the paramagnetic phase of Mn$_{1-x}$Fe$_{x}$Si solid solutions}
\date{\today}

\author{S.~{}V.~{}Demishev}
\affiliation {Prokhorov General Physics Institute of RAS, 38 Vavilov str.,  119991 Moscow, Russia} 
\author{I.~{}I.~{}Lobanova}
\author{V.~{}V.~{}Glushkov}
\affiliation {Prokhorov General Physics Institute of RAS, 38 Vavilov str.,  119991 Moscow, Russia} \affiliation{Moscow Institute of Physics and Technology , 9 Institutskiy per., 141700 Dolgoprudny, Moscow region, Russia}

%\altaffiliation[]{Moscow Institute of Physics and Technology , 9 Institutskiy per., 141700 Dolgoprudny, Moscow region, Russia}
\author{V.~{}Yu.~{}Ivanov}
\author{T.~{}V.~{}Ischenko}
\author{V.~{}N.~{}Krasnorussky}
\author{N.~{}E.~{}Sluchanko}
\affiliation {Prokhorov General Physics Institute of RAS, 38 Vavilov str.,  119991 Moscow, Russia} 
%\affiliation{Moscow Institute of Physics and Technology , 9 Institutskiy per., 141700 Dolgoprudny, Moscow region, Russia}
\author{N.~{}M.~{}Potapova}
\author{V.~{}A.~{}Dyadkin}
\author{S.~{}V.~{}Grigoriev}
\affiliation{Petersburg Nuclear Physics Institute, Gatchina, 188300 Saint-Petersburg, Russia}

\begin{abstract}
 The magnetization field and temperature dependences in the paramagnetic phase of Mn$_{1-x}$Fe$_x$Si solid solutions with $x<0.3$ are investigated in the range $B<5$~T and $T<60$~K. It is found that field dependences of the magnetization $M(B,T=const)$ exhibit scaling behaviour of the form $B\partial M/\partial B-M=F(B/(T-T_s))$, where $T_s$ denotes an empirically determined temperature of the transition into the magnetic phase with fluctuation driven short-range magnetic order and $F(\xi)$ is a universal scaling function for given composition. The scaling relation allowed concluding that the magnetization in the paramagnetic phase of Mn$_{1-x}$Fe$_x$Si is represented by the sum of two terms. The first term is saturated by the scaling variable $\xi=B/(T-T_s)$, whereas the second is linearly dependent on the magnetic field. A simple analytical formula describing the magnetization is derived and applied to estimates of the parameters characterizing localized magnetic moments in the studied system. The obtained data may be qualitatively interpreted assuming magnetic inhomogeneity of the paramagnetic phase on the nanoscale. 
 \end{abstract}

% insert suggested PACS numbers in braces on next line
% insert suggested keywords - APS authors don't need to do this
%\keywords{}

\maketitle

\section{Introduction}
Spiral magnets based on manganese monosilicide, MnSi, continue attracting attention mainly due to various aspects of skyrmion physics\cite{Pappas09,Pappas11,Neubauer09,Tonomura12,Li13}. As long as skyrmions are expected developing in the so-called A-phase, this region on the magnetic phase diagram is the focus of researchers, whereas another magnetic phases are often considered as more simple and less interesting. From the general point of view, the simplest magnetic phase is the paramagnetic (PM) phase where strong correlation effects are not expected and which is believed to be well described by Moriya theory of spin fluctuations\cite{Moriya85}. However, some anomalies of the magnetic and transport properties of the PM phase in MnSi have been reported
 recently\cite{Storchak11,Demishev11,Demishev12}. For example, the formation of spin-polaron states of the ferron type at temperatures well above the Curie temperature $T_C$ was suggested for the explanation of the $\mu$SR experiments\cite{Storchak11}. The analysis of the transport and magnetic resonance data also favors the explanation of magnetic properties of MnSi by spin polarons rather than within standard itinerant magnetism approach\cite{Demishev11,Demishev12}. Thus, the physical idea, combining the results of the above works\cite{Storchak11,Demishev11,Demishev12}, is the concept of magnetic inhomogeneity on the nanometer scale, which should be manifested in the magnetic properties of manganese monosilicide.

In this connection, the study of Mn$_{1-x}$Fe$_x$Si solid solutions may be 
prospective for better
 understanding of the magnetic inhomogeneities problem. According to
 Refs.~\onlinecite{Nishihara84,Grigoriev09,Bauer10} the increase of the iron content leads to the suppression of the transition into the helical phase with long-range magnetic order and formation of the quantum critical point at $x=x^*\approx 0.12-0.15$ for which $T_C(x^*)=0$
 \cite{Nishihara84,Grigoriev09,Bauer10}. Some experimental\cite{Bauer10,Grigoriev11} and theoretical works\cite{Kruger12,Tewari06} suggests that an intermediate spiral-based phase with short-range magnetic order may be formed either in the vicinity of $T_C(x)$ or in the range $x>x^*$ due to strong spiral fluctuations extending into the paramagnetic phase\cite{Pappas09,Pappas11,Bauer10,Grigoriev11}. Consequently certain regions of the magnetic phase diagram of Mn$_{1-x}$Fe$_x$Si solid solutions may correspond to the formation of spatially inhomogeneous magnetic state with specific magnetization.

Magnetization and magnetic susceptibility studies are often aimed at obtaining information about phase boundaries on the magnetic phase diagrams of Mn$_{1-x}$Fe$_x$Si\cite{Grigoriev11,Bauer10,Grigoriev11,Thessieu97}. Therefore the regions, where the magnetization field and temperature dependences $M(B,T)$ undergo abrupt changes due to magnetic transitions appears in the focus of interest, whereas the paramagnetic phase characterized by smooth $M(B,T)$ curves attracts less attention. 

In the present work the magnetization in the paramagnetic phase of Mn$_{1-x}$Fe$_x$Si solid solutions with $x<0.3$ is investigated in the range $B<6$~T and $T<60$~K. We show that $M(B,T)$ function in the studied system may be represented as a sum of two terms, one of which shows universal scaling behavior. A simple analytical expression describing experimental $M(B,T)$ data in the studied $B-T$ domain is suggested. The analysis of the approximation parameters concentration dependences indicates possible magnetic inhomogeneity of the paramagnetic phase on the nanoscale. 

\section{Experimental details}
Single crystals of Mn$_{1-x}$Fe$_x$Si solid solutions were synthesized by both Czochralski and Bridgeman methods. The crystal structure of the samples was controlled by the X-ray Laue diffraction. EPMA was applied for determination of the samples chemical composition. Assuming formula (Mn$_{1-x}$Fe$_x$)$_{1+y}$Si$_{1-y}$ we found that stoichiometry of crystals was kept at the level $y\sim 0.01-0.005$ comparable with the absolute error of our EPMA measurements. The parameters $x$ characterizing studied samples de facto were found to deviate from the nominal ones calculated for the initial ingot and the discrepancy could be as big as $\sim 0.05$. Below we will present only real iron concentration in the samples. The third digit of the $x$ number is used as a reference and corresponds to average Fe content obtained by several scans along the sample surface. The magnetization and magnetic susceptibility data in the magnetic fields up to 5~T for temperatures in the range 1.8-60~K were obtained with the help of SQUID magnetometer (Quantum Design).
\section{Experimental data and magnetization scaling in the paramagnetic phase of M\lowercase{n}$_{1-x}$F\lowercase{e}$_x$S\lowercase{i}}
Before discussing magnetic properties in the paramagnetic phase it worth considering location of the paramagnetic phase on the $T-x$ magnetic phase diagram. At present, it is widely accepted\cite{Bauer10,Grigoriev11} that in Mn$_{1-x}$Fe$_x$Si the transition into spiral phase with long-range magnetic order at $T_C$ may be preceded by formation of a fluctuation driven intermediate short-range ordered magnetic phase with the transition temperature $T_s>T_C$. These empirical findings \cite{Bauer10,Grigoriev11} are supported by theoretical analysis\cite{Kruger12,Tewari06}. Nevertheless at a moment it is not clear whether transition temperature $T_s$ is a true sharp phase boundary or this characteristic temperature merely marks a crossover region where spiral fluctuations in Mn$_{1-x}$Fe$_x$Si system slow down and freeze. Consequently for $x<x^*$ the region of the paramagnetic phase may correspond either to $T>T_s(x)$ or to $T>T_C(x)$. As long as diapason $x>x^*$ is characterized by the absence of the long-range magnetic order\cite{Nishihara84,Grigoriev09,Bauer10},  the area of the paramagnetic phase may be  defined as $T>T_s(x)$ or $T>0$ in this concentration range. 

\begin{figure}
\includegraphics[width=1\linewidth]{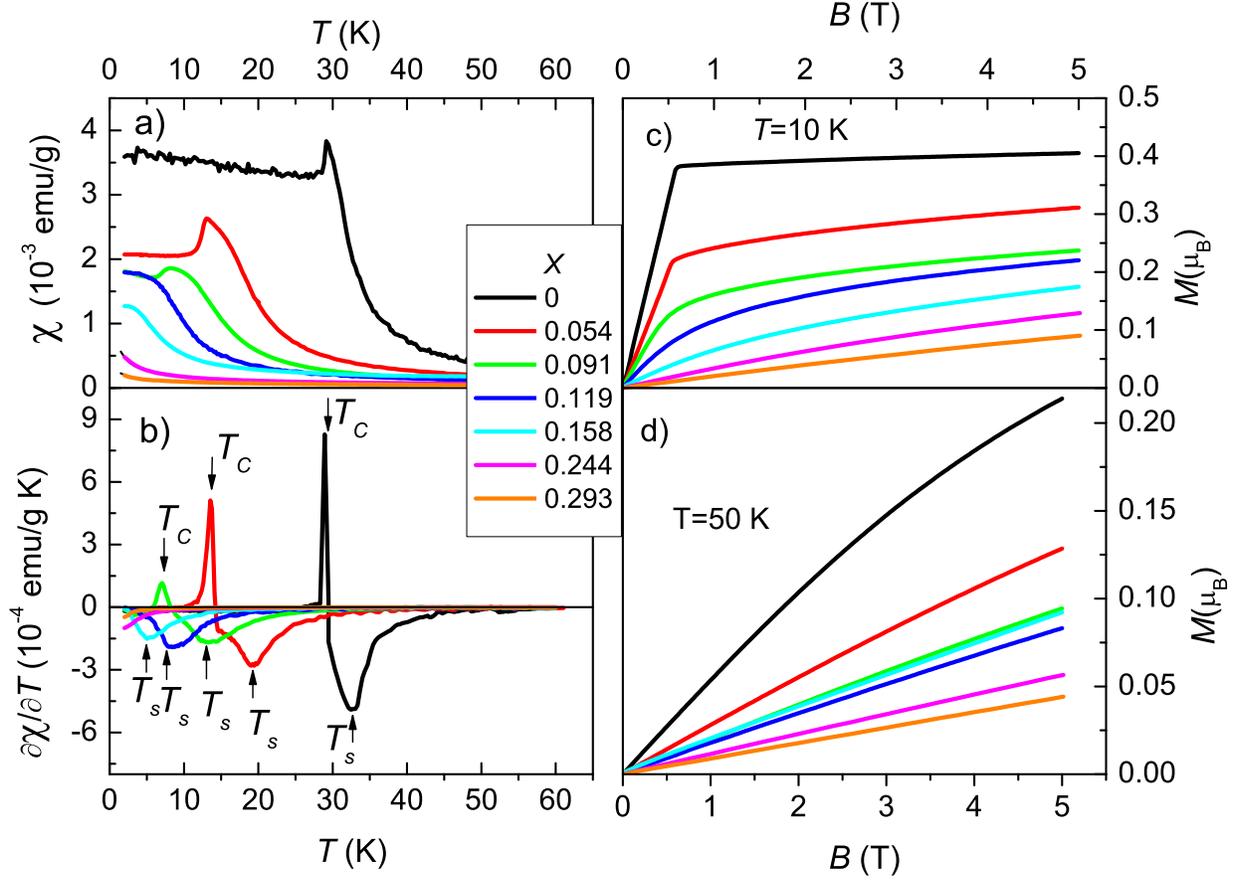}
\caption{\label{Fig1}(Color on-line) Magnetic susceptibility (a), temperature derivative of magnetic susceptibility (b) and magnetization at $T=10$~K (c) and $T=50$~K (d) in Mn$_{1-x}$Fe$_x$Si solid solutions. Corresponding iron concentrations are denoted by color as shown in the legend.}
\end{figure}

Detailed comparison of polarized neutron scattering data and temperature dependences of magnetic susceptibility $\chi(T)$ in Mn$_{1-x}$Fe$_x$Si system was carried out in\cite{Grigoriev11}. The analysis of the data concerning magnetic structure at various temperatures allowed concluding that transitions into the phases with short-range and long-range magnetic orders unambiguously correspond to the inflection points of the $\chi(T)$ curve\cite{Grigoriev11}. For example, the magnetic susceptibility peak in Fig.~\ref{Fig1},a does not occur at $T_C(x)$ as would be naive to assume, and real transition temperature into spiral phase with long-range magnetic order is located below mentioned peculiarity. Therefore for determination of the magnetic transition temperatures it is instructive to analyze temperature dependences of the magnetic susceptibility derivatives $\partial\chi/\partial T=f(T)$ (Fig.~\ref{Fig1},b). According to Ref.~\onlinecite{Grigoriev11} the broad minima of the $\partial\chi/\partial T$ should mark transition into fluctuation driven spiral phase with short-range magnetic order at $T_s(x)$, whereas narrow maximum of the magnetic susceptibility derivative denotes formation of the magnetic phase with long-range spiral magnetic order at $T_C(x)$.

The experimental temperature dependences of the magnetic susceptibility and its derivative for Mn$_{1-x}$Fe$_x$Si are shown in Fig.~\ref{Fig1},a,b. It is visible that for any $x$ the inequality $T_s(x)>T_C(x)$ is valid and increase in iron concentration suppresses magnetic transition temperatures. The transition into magnetic phase with long-range order may be detected for $x<x^*\approx 0.11$ only. In agreement with the previous experimental results and theoretical models\cite{Grigoriev09,Bauer10,Kruger12,Tewari06} it is reasonable to assume from the data in Fig.~\ref{Fig1} that $T_C(x>x^*\approx 0.11)=0$.  However, the transition into the phase with short-range magnetic order lasts up to $x\sim 0.24$, i.e. this phase exists even for $x>x^*$ (Fig.~\ref{Fig1},b). For iron concentrations exceeding the latter value the transition temperature $T_s(x)$ is not observed, which means that this parameter is either less than the lowest temperature used in our experiments or turns to zero. 

The examples of raw magnetization data for Mn$_{1-x}$Fe$_x$Si are presented in Fig.~\ref{Fig1},c,d. The area of the paramagnetic phase corresponds to smooth $M(B)$ curves, whereas those demonstrating a kink belongs to magnetically ordered phase and this type of the magnetization field dependences will be excluded from further analysis.

When discussing possible scaling for magnetization we suppose that it can be presented in the mathematical form $M(B,T)=M(\xi)$ for fixed $x$, where the scaling variable $\xi$ is given by $\xi=B/\Theta(T)$. The above consideration of the paramagnetic phase boundary location allows suggesting several forms for the function $\Theta(T)$, which may be tried for description of the $M(B,T)$ data. Namely, it is reasonable examine the cases $\Theta(T)=T$, $\Theta(T)=T-T_C$ and $\Theta(T)=T-T_s$ which may be valid in different concentration ranges. At the first step, an assumption that magnetization itself scales if the scaling variable is chosen properly may be considered.

\begin{figure}[t]
\includegraphics[width=1\linewidth]{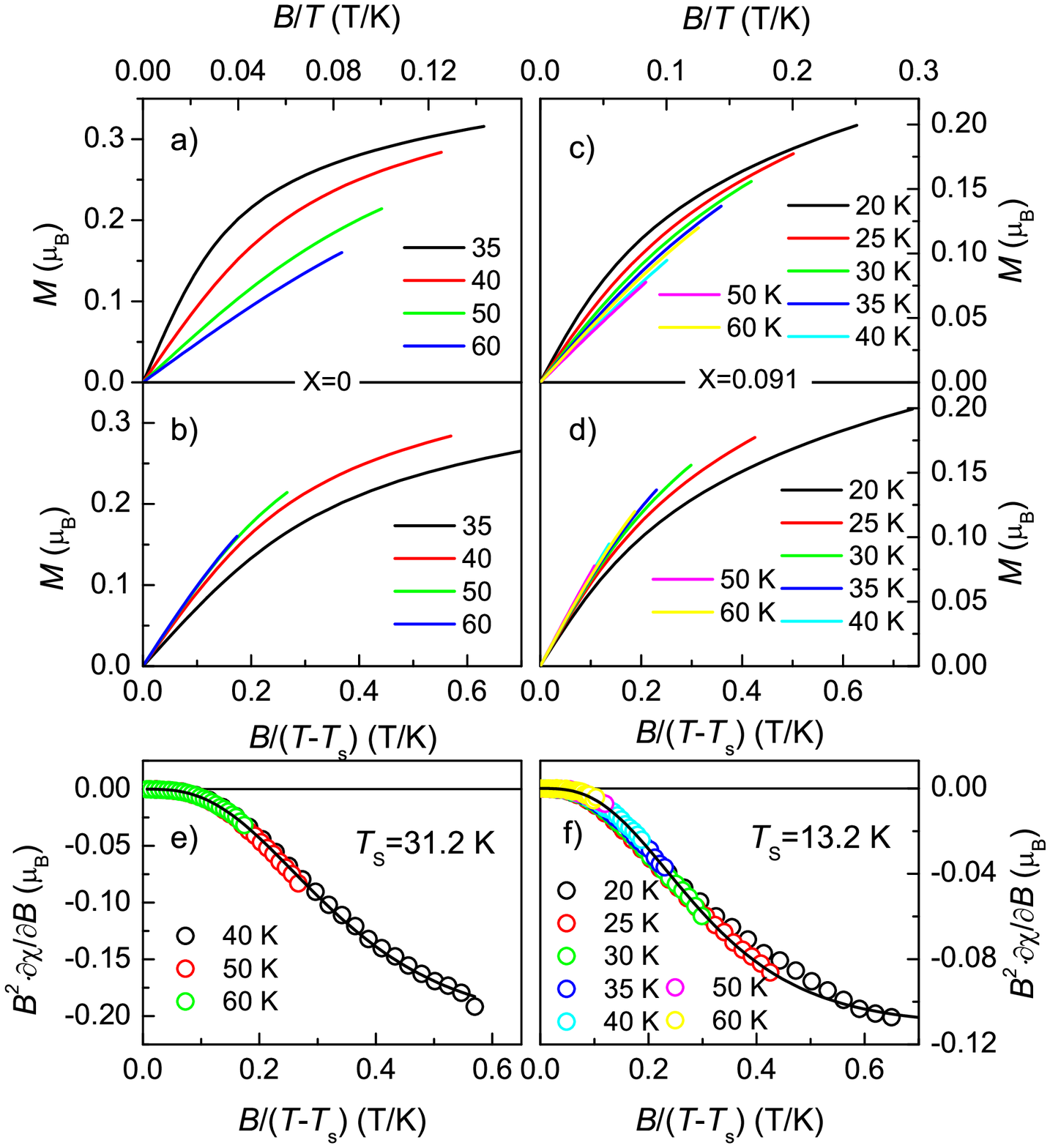}
\caption{\label{Fig2}(Color on-line) Field dependences of magnetization in units of Bohr magneton per Mn site for $x=0$ (a,b,e) and $x=0.091$(c,d,f) in the cases of different scaling variables: $\xi=B/T$ (a,c) and $\xi=B/(T-T_s)$ (b,d). Panels e, f  represent scaling function in equation (2): points - experiment, solid lines - model fit by equation (5) (see text for details).}
\end{figure}

\begin{figure}[t]
\includegraphics[width=1\linewidth]{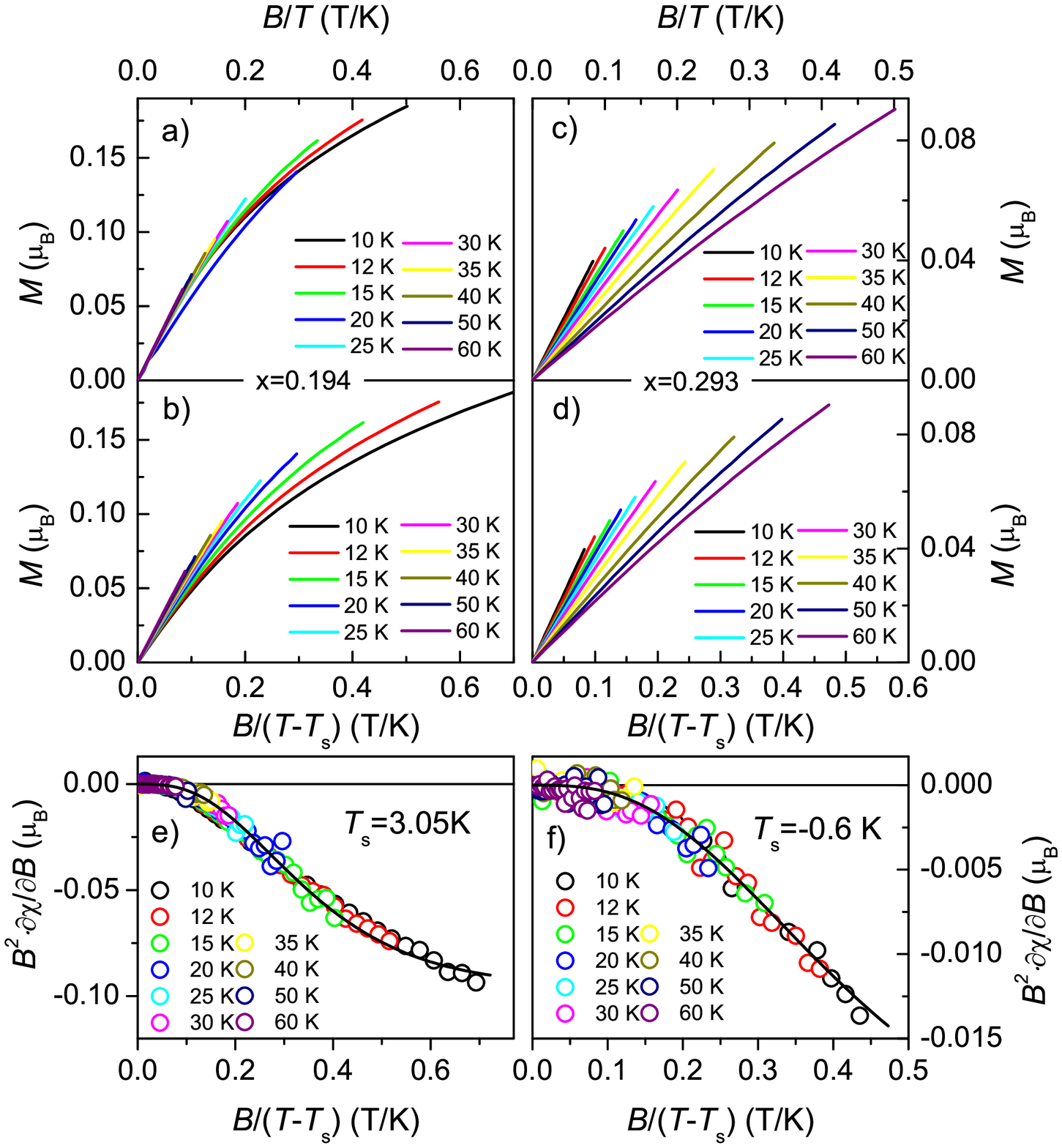}
\caption{\label{Fig3} (Color on-line) Field dependences of magnetization in units of Bohr magneton per Mn site for $x=0.194$ (a,b,e) and $x=0.293$ (c,d,f) in the cases of different scaling variables:$\xi=B/T$ (a,c) and $\xi=B/(T-T_s)$ (b,d). Panels e, f  represent scaling function in equation (2): points - experiment, solid lines - model fit by equation (5) (see text for details).}
\end{figure}

Analysis of the experimental $M(B,T)$ data show that any of the above forms for $\xi=B/\Theta(T)$ does not allow obtaining magnetization scaling for any $x$. The examples for the scaling variables $\xi=B/T$ and $=B/(T-T_s)$ are shown in Figs.~\ref{Fig2},\ref{Fig3}, panels a,c and b,d respectively. The choice $\xi=B/(T-T_C)$ provides results similar to that for $\xi=B/(T-T_s)$ as long as the characteristic temperatures $T_C$ and $T_s$ are relatively close in the diapason $x<x^*$.

The above consideration may lead to two possible conclusions. The supposition that there is no magnetization scaling in Mn$_{1-x}$Fe$_x$Si system at all constitutes the first opportunity. The second assumption is that magnetization of Mn$_{1-x}$Fe$_x$Si consists of several contributions, one of which may be scaled with $\xi=B/\Theta(T)$, whereas another terms may depend on the variables different from $\xi$, thus masking the magnetic contribution, which may possess scaling behavior. A hint of the possible solution of this problem may be found in the field dependence of magnetization in the phase with long-range magnetic order. It is known that in MnSi for $T<T_C$ magnetization does not saturate and after initial rapid growth of $M(B)$ with a magnetic field there is still slow linear increase of magnetization \cite{Nishihara84,Demishev12,Bauer10}. The same behavior corresponds to the magnetically ordered phase at $T<T_C$ in Mn$_{1-x}$Fe$_x$Si solid solutions with $x<x^*$ (Fig.~\ref{Fig1},c,d). Moreover, the pulsed field measurements up to $B\sim 30$~T unambiguously demonstrated the presence of the linear term in magnetization not only in the spiral phase but also in the paramagnetic phase of MnSi\cite{Sakakibara82}. Basing on observations, we assume that in the paramagnetic phase the magnetization of Mn$_{1-x}$Fe$_x$Si has the structure
\begin{equation}
M(B,T)=M_0\cdot\phi(\xi)+A(T)\cdot B.
\end{equation}
In Eq.~(1) function $\phi(\xi)$ satisfy conditions $\phi(\xi\ll 1)\sim\xi$ and $\phi(\xi\gg 1)=1$, argument $\xi=B/\Theta(T)$ is one of the possible scaling variables described above, $M_0$ denotes saturated magnetization for the first term and function $A(T)$ possesses temperature dependence different from $\phi(B/\Theta(T))$. In other words, our hypothesis consists in the statement that there is an additional linear term in magnetization disguising expected scaling behavior.

Defining magnetic susceptibility as $\chi=M(B,T)/B$ it is possible to transform Eq.~(1):
\begin{equation}
B^2\frac{\partial\chi}{\partial B}\equiv B\frac{\partial M}{\partial B}-M=F(\xi).
\end{equation}
Eq.~(2) is valid for any form of $A(T)$ and scaling function $F(\xi)$ is given by
\begin{equation}
F(\xi)=M_0\left[\xi\phi'(\xi)-\phi(\xi)\right].
\end{equation}
This transformation allows eliminating any linear term in magnetization (including linear term in $\phi(\xi)$) and therefore rendering the field dependences of magnetization $M(B,T=const)$ of the coordinates (2) it is possible to reveal unmasked scaling behavior if any.

The aforementioned analysis was performed for various forms of scaling variable $\xi$. It if found that for $\xi=B/(T-T_s)$, where $T_s$ is defined from experiment as shown in Fig.~\ref{Fig1},b, the magnetization $M(B,T=const)$ converted in coordinates (2) demonstrate scaling behavior for all studied samples of Mn$_{1-x}$Fe$_x$Si solid solutions with $x<x_c\sim 0.24$ (Figs.~\ref{Fig2},\ref{Fig3}, panels e-f)). Due to the absence of $\partial\chi/\partial T$ minima for $x>0.24$ the characteristic temperature $T_s$ in this concentration range was considered as an additional fitting parameter. The best results were obtained for small (less that 1~K) negative value of $T_s$ (see Fig.~\ref{Fig3},f). 

\section{Analysis of two magnetic contributions}
 
Observation of scaling behavior in coordinates $B^2\partial\chi\partial B=F(B/(T-T_s))$ confirms hypothesis that magnetization structure may be expressed by Eq.~(1). However, mathematical transformation discussed above violates the equivalence of the transition from Eq.~(1) to Eq.~(2) and it is not possible to use experimentally obtained scaling functions $F(B/(T-T_s))$ (Figs.~\ref{Fig2},\ref{Fig3}, panels e-f) for direct separation of the linear and saturating terms in magnetization. Therefore, in order to estimate different magnetic contributions in Eq.~(1) it is reasonable to use the model expression for $\phi(\xi)$. Namely we assume that
\begin{equation}
\phi(\xi)=\tanh\left(\frac{\mu^*\xi}{k_B}\right),
\end{equation}
and accordingly 
\begin{equation}
F(\xi)=M_0\left(-\tanh\left(\frac{\mu^*\xi}{k_B}\right)+\frac{\frac{\mu^*\xi}{k_B}}{\cosh\left(\frac{\mu^*\xi}{k_B}\right)}\right).
\end{equation}

It is visible from Figs.~\ref{Fig2},\ref{Fig3} that model expression for scaling function (5) provides a satisfactory approximation of experimental data (solid lines in panels e-f) with the help of two fitting parameters $M_0$ and $\mu^*$. Hereafter the latter parameter is denoted as an effective magnetic moment and its possible physical meaning will be discussed in a subsequent section. It is worth noting that both values of $M_0$ and $\mu^*$ depend only on iron concentration $x$ and does not depend on temperature.

\begin{figure}
\includegraphics[width=1\linewidth]{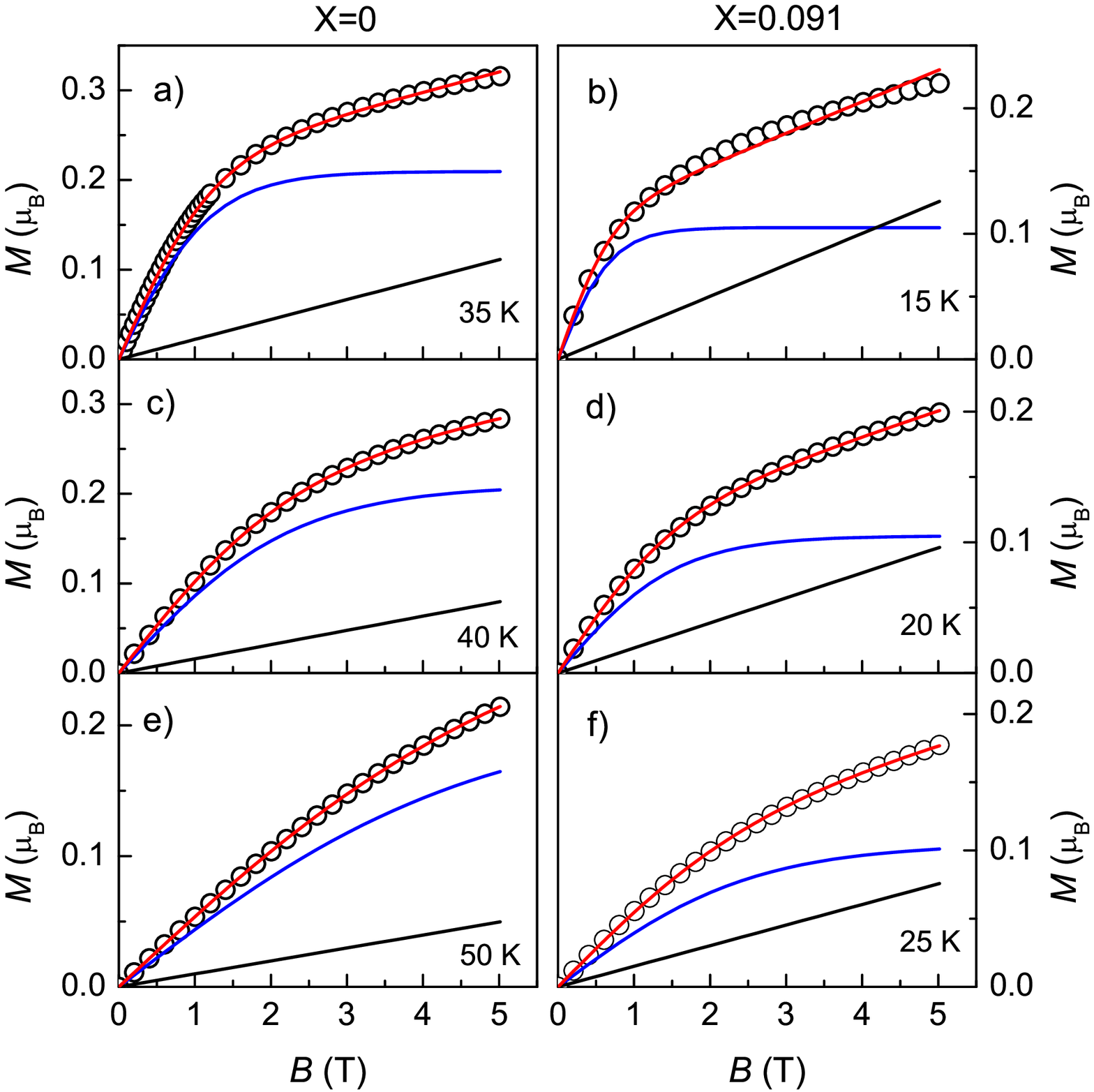}
\caption{\label{Fig4}(Color on-line) Field dependences of magnetization for $x=0$ (a, c, e) and $x=0.091$ (b,d,f) at various temperatures: 35~K (a), 15~K (b), 40~K (c), 20~K (d), 50~K (e) and 25~K (f). Points and solid lines represent experimental data and model approximation with the help of Eq.~(6) accordingly. Saturating and linear contributions to magnetization are shown by red and blue lines respectively. The magnetization is given in units of Bohr magneton per Mn site.}
\end{figure}

\begin{figure}
\includegraphics[width=1\linewidth]{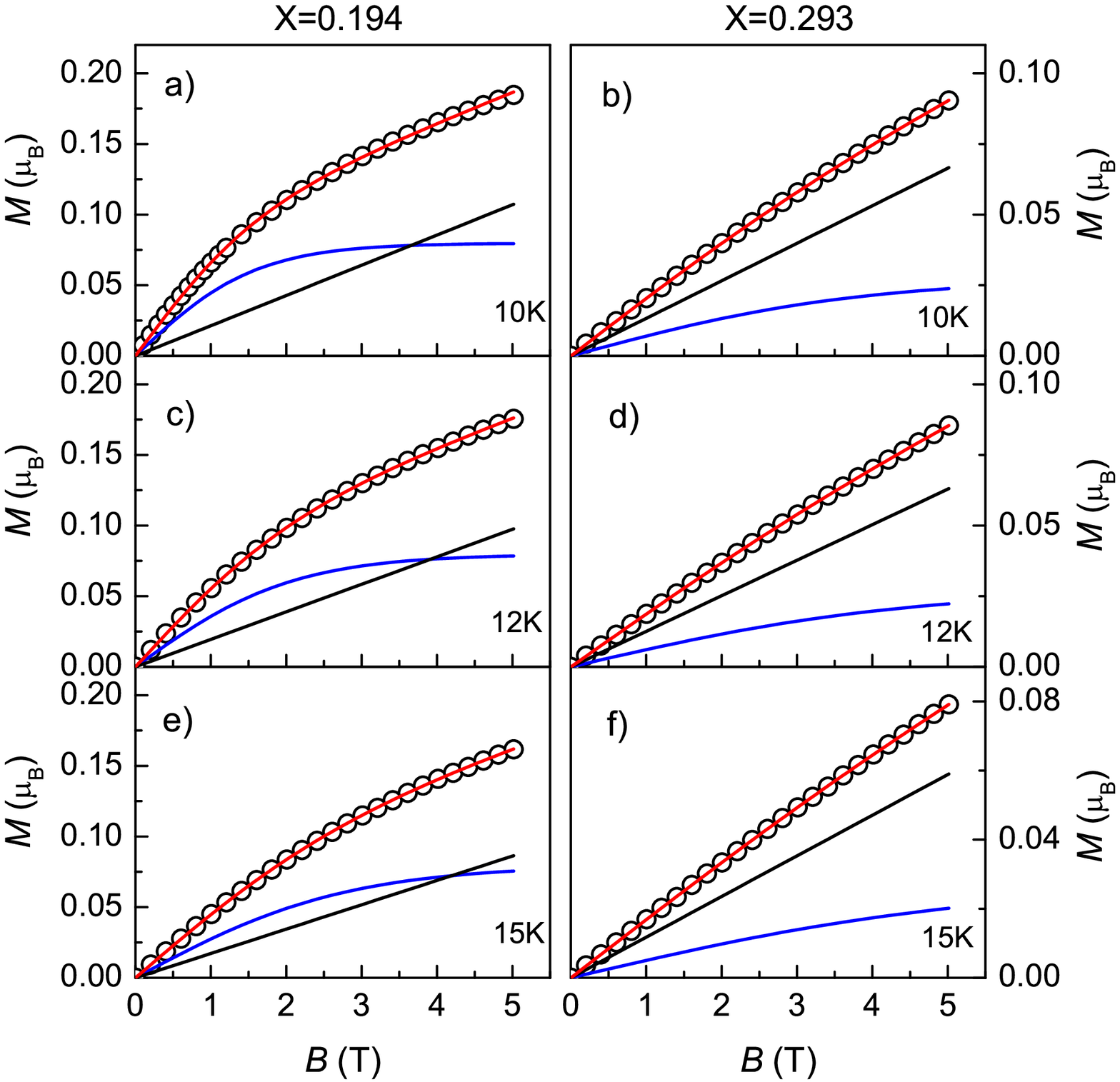}
\caption{\label{Fig5}(Color on-line) Field dependences of magnetization for $x=0.194$ (a, c, e) and $x=0.293$ (b,d,f) at various temperatures: 35~K (a), 15~K (b), 40~K (c), 20~K (d), 50~K (e) and 25~K (f). Points and solid lines represent experimental data and model approximation with the help of Eq.~(6) accordingly. Saturating and linear contributions to magnetization are shown by red and blue lines respectively. The magnetization is given in units of Bohr magneton per Mn site.}
\end{figure}

Fitting of the scaling function $F(B/(T-T_s))$ does not allow finding coefficient $A(T)$ in Eq.~(1) and hence it is not possible to estimate correctly the errors in determination of $M_0$ and $\mu^*$ although chosen approximation for $\phi(\xi)$ (Eq.~(4)) seems grounded enough. For that reason, the following form for analysis of the field dependences of magnetization was examined
\begin{equation}
M(B,T=const)=M_0\tanh(m\cdot B)+A\cdot B.
\end{equation}
This expression contains three fitting parameters $M_0$, $m$ and $A$, which may be functions of temperature for each fixed iron concentration. Examples of magnetization data analysis for Mn$_{1-x}$Fe$_x$Si with the help of Eq.~(6) at different temperatures are presented in Figs.~\ref{Fig4},\ref{Fig5}. It is visible that the model form (6) adequately describes the field dependences $M(B,T=const)$ for various $x$. The magnitudes of the linear term and saturating terms in (6) are comparable for $B\sim 5$~T. Moreover, for highest iron concentration the linear contribution to magnetization becomes bigger than the saturating part (Fig.~\ref{Fig5},b,d,f).


\begin{figure}
\includegraphics[width=1\linewidth]{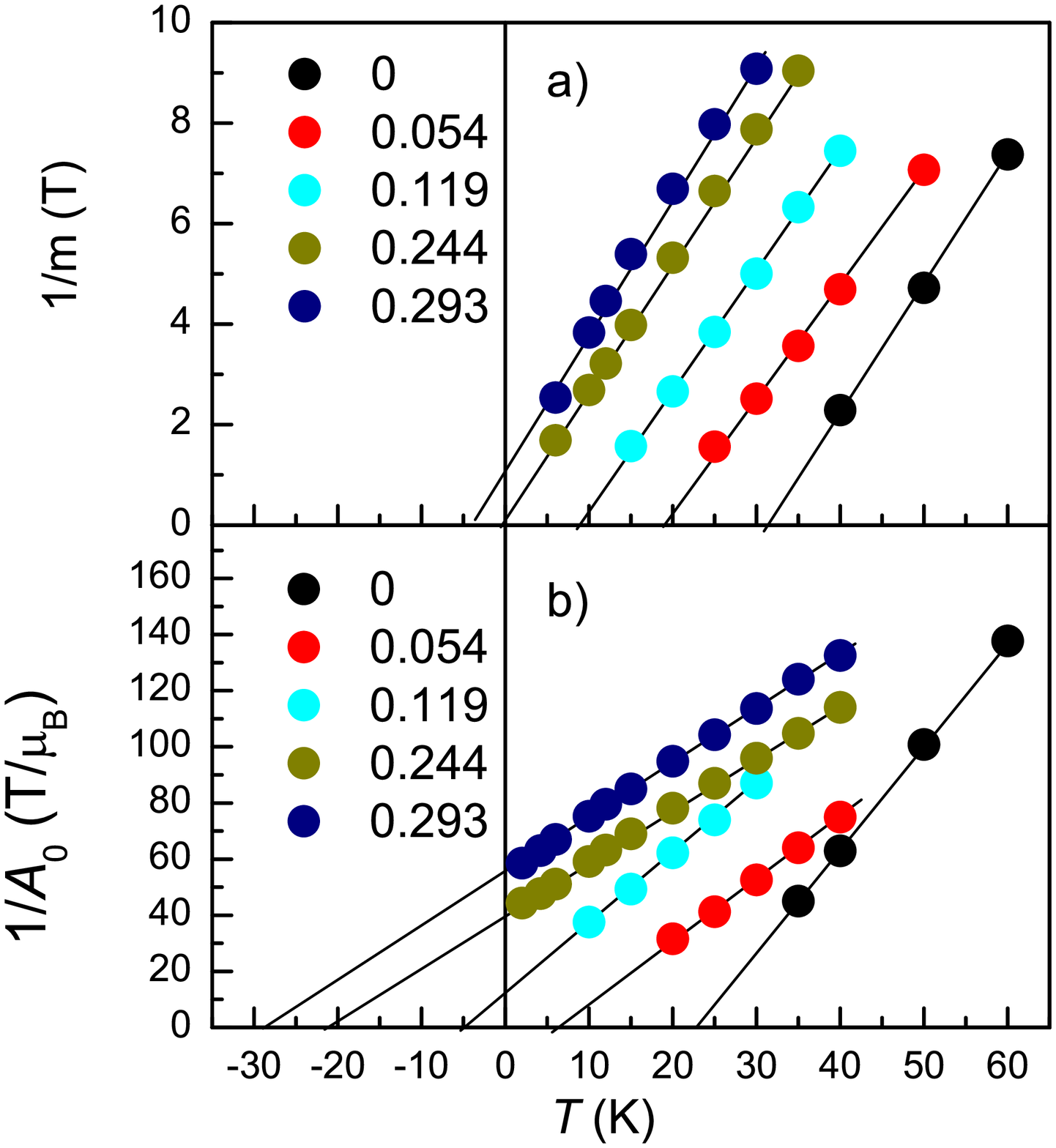}
\caption{\label{Fig6}(Color on-line) Temperature dependences $m^{-1}=f(T)$ (a) and $A_0^{-1}=f(T)$ (b) (see text for details).}
\end{figure}

\begin{figure}
\includegraphics[width=1\linewidth]{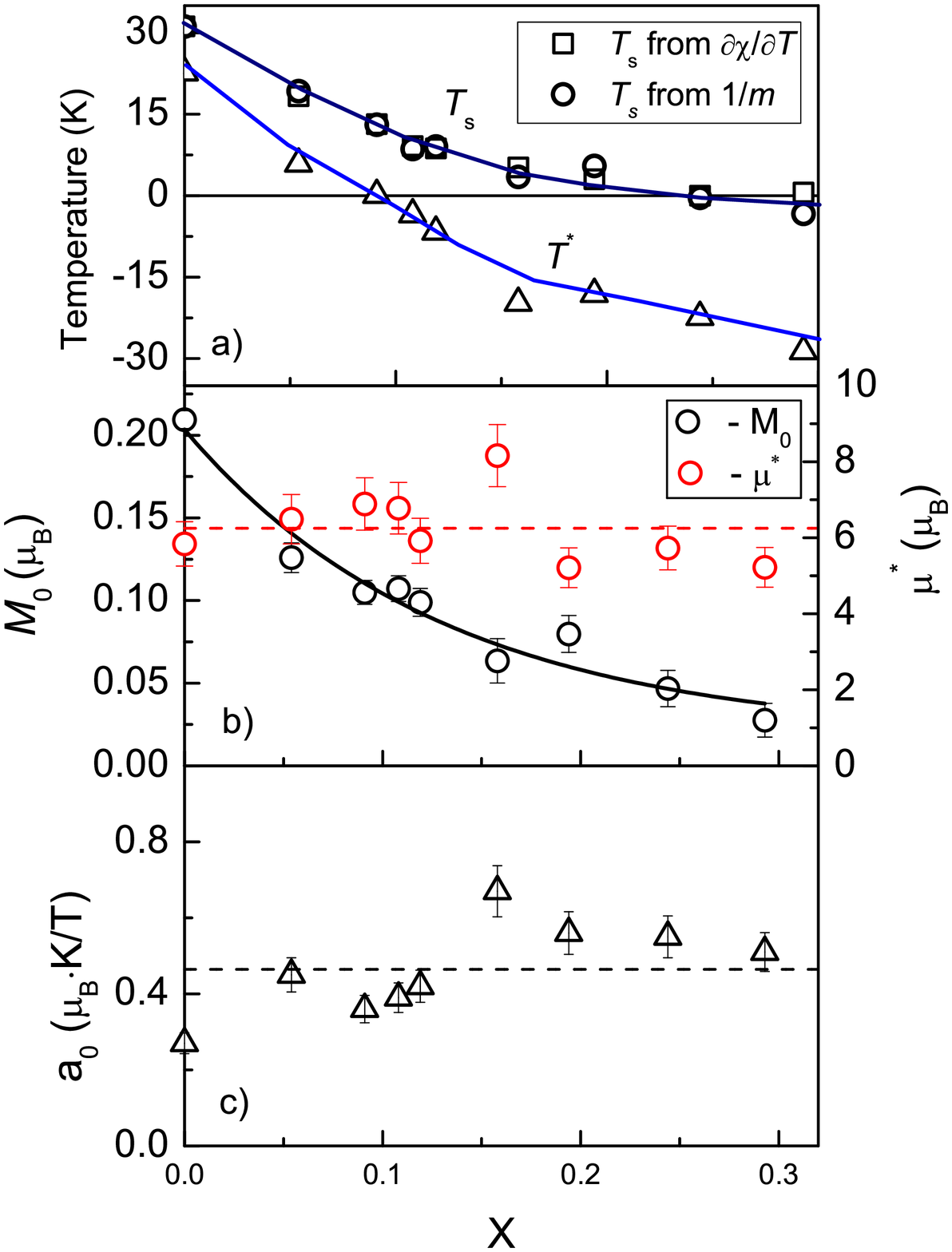}
\caption{\label{Fig7}(Color on-line) Concentration dependences of various parameters in Eq.~(8): characteristic temperatures $T_s$ and $T^*$ (a), $M_0$ and $\mu^*$ (b) and $a_0$ (c). Solid lines are guides for eyes, dashed lines correspond to averaged values of $\mu^*$ and $a_0$. In the panel (b) parameter $M_0$ is given in units of Bohr magneton per Mn site and parameter $\mu^*$ is given in the units of Bohr magneton.}
\end{figure}

The approximation procedure with the help of Eq.~(6) demonstrated good convergence and stability. It is found that in the studied case the parameter $M_0$ does not depend on temperature in agreement with the results provided by scaling function $F(B/(T-T_s))$. Moreover, both methods for the finding of the saturated magnetization $M_0$ provide practically coinciding values of this quantity. In order to analyze $m(T)$ data the obtained temperature dependences were plotted in coordinates $m^{-1}=f(T)$ as long as the above consideration suggests relation $m^{-1}\sim(T-T_s)$. The results are shown in Fig.~\ref{Fig6},a. Good linear dependence allows finding unambiguously the characteristic temperature $T_s$ by extrapolation of the $m^{-1}(T)$ lines to the value $m^{-1}=0$. The obtained $T_s(x)$ dependence can be compared with experimental one subtracted from the $\partial\chi/\partial T=f(T)$ data shown in Fig.~\ref{Fig1},b. It can be concluded from Fig.~\ref{Fig7},a that calculated and experimental functions $T_s(x)$ agrees very well for $x<x_c\sim 0.24$. This may serve as an additional argument in favor for correctness of two magnetic contributions separation shown in Figs.~\ref{Fig4},\ref{Fig5}. For $x>0.24$ the minima on the $\partial\chi/\partial T=f(T)$ curves are not observed and $m^{-1}(T)$ data (Fig.~\ref{Fig6},a) corresponds to $T_s\approx 0$ $(x=0.244)$ and $T_s\approx-0.6$~K (Fig.~\ref{Fig7},a). Comparing this finding with the consequence of scaling function analysis it is possible to expect that $T_s$ may change sign in diapason $x>x_c\sim 0.24$.

The similar consideration was carried out in the case of the $A(T)$ temperature dependences obtained from fitting of the $M(B,T=const)$ curves (Figs.~\ref{Fig4},\ref{Fig5}). The data shown in Fig.~\ref{Fig6},b suggest that this coefficient may be presented in the form
\begin{equation}
A(T)=\frac{a_0}{T-T^*},
\end{equation}
where temperature $T^*\neq T_s$ strongly depends on $x$ changing sign in the vicinity of $x^*\sim 0.11$ (Figs.~\ref{Fig6},b and \ref{Fig7},a). Thus in Mn$_{1-x}$Fe$_x$Si two terms of magnetization (Eq.~(1)) are described by two characteristic temperatures differ from one another and showing specific concentration dependences.

The effective magnetic moment $\mu^*$ and parameter $a_0$ in Eq.~(7) were calculated from the slopes $\partial m^{-1}/\partial T =k_B/\mu^*$ and $\partial A^{-1}/\partial T =1/a_0$ of the corresponding lines in Fig.~\ref{Fig6},a,b. The resulting concentration dependences of these quantities are shown in Fig.~\ref{Fig7},b and Fig.~\ref{Fig7},c respectively.  We wish to mark that these two parameters vary weakly around average values $\mu^*\sim 6$ $\mu_B$ and $a_0\sim 0.4$~$\mu_B$K/T (Fig.~~\ref{Fig7},b,c). At the same time (Fig.~\ref{Fig7}\,b) the saturated magnetization per Mn site decreases with $x$ in the studied diapason by more that 8 times from $M_0\sim 0.21~\mu_B (x=0)$ to $M_0\sim 0.025~\mu_B (x=0.293)$. The $M_0$ and $\mu^*$ values obtained from the scaling function analysis lie within error bars shown in Fig.~\ref{Fig7},b.

\section{Discussion}
Analysis of the magnetization data in Mn$_{1-x}$Fe$_x$Si shows that $M(B,T)$ function in this system acquires the form (1), which for $x<0.3$ and $B<5$~T can be well approximated as
\begin{equation}
M(B,T)=M_0\tanh\left(\frac{\mu^*B}{k_B(T-T_s)}\right)+\frac{a_0}{T-T^*}B.
\end{equation}
The concentration dependences of the parameters found in the present work suggest that $\mu^*(x)\approx const$ and $a_0(x)\approx const$, whereas $M_0(x)$, $T_s(x)$ and $T^*(x)$ strongly depend on iron content (Fig.~\ref{Fig7},a-c). It is worth noting that none of the above parameters is a function of temperature and temperature dependence of magnetization in Mn$_{1-x}$Fe$_x$Si may be completely described by Eq.~(8). 

As a summary of experimental facts, the empirical formula (8) leads to several consequences important in practical and theoretical sense. First of all, it is necessary to point out that temperature dependence of magnetic susceptibility at low temperatures ($T<60$~K) in Mn$_{1-x}$Fe$_x$Si system will always deviate from Curie-Weiss law owing to inequality of $T_s$ and $T^*$. Therefore any estimates of effective on-site magnetic moment magnitude and (or) concentration of the magnetic dipoles based on Curie-Weiss analysis could be misleading for the case studied. At the same time, Eq.~(8) may be applied to model calculations of magnetocaloric effect in Mn$_{1-x}$Fe$_x$Si essential, for example, in the case of pulsed magnetic field measurements\cite{Sakakibara82}.

Secondly, assuming that classical itinerant picture of MnSi magnetism\cite{Moriya85} is valid, it is not possible to suggest any physical mechanism for existence of two magnetic contributions. Therefore, if itinerant ansatz is taken for granted, Eqs.~(1) and (8) are nothing but some mathematical approximations of experimental data, which does not have direct physical sense. However, even in the considered approach, the singularity of denominator at $T_s$ in the first term of (8) suggests that the transition between paramagnetic phase and phase with short-range magnetic order\cite{Grigoriev11} occurs at sharp boundary and can not be treated as a simple crossover phenomenon.

In addition, it is necessary to mark that any of the magnetic contributions in Mn$_{1-x}$Fe$_x$Si obtained in the present work could not be associated with the effect of substitution of manganese with iron as long as both terms are present in pure MnSi. Moreover, the total magnetic moment of the sample decreases with $x$ (Figs.~\ref{Fig4},\ref{Fig5}), and therefore it is reasonable to suppose that the observed phenomena including magnetization scaling are due to Mn magnetic subsystem. 

In order to explain the observed magnetization structure (Eq.~(8)) it is possible to consider an alternative description of Mn$_{1-x}$Fe$_x$Si magnetism based on Heisenberg-type localized magnetic moments. At present this type of models for MnSi magnetism are often considered as a simplified approximation of itinerant case. However, the experimental study of magnetic resonance and magnetoresistance\cite{Demishev11,Demishev12} and LDA calculations\cite{Corti07} indicate real existence of localized magnetic moments (LMM) on Mn sites.  In order to interpret the reduction of the saturated magnetic moment\cite{Moriya85}, realization of Yosida mechanism of magnetic scattering\cite{Demishev12}, presence of strong spin fluctuations in neutron scattering\cite{Ishikawa77} and electron spin resonance\cite{Demishev11,Demishev12} data in the Heisenberg paradigm, it is necessary to imply specific mechanism of screening of localized magnetic moments\cite{Demishev11,Demishev12}. According to the hypothesis formulated in Ref.~\onlinecite{Demishev11,Demishev12} itinerant electrons form quasibound state in the vicinity of the Mn ion. In this state spin of the electron tends to be oriented opposite with respect to LMM on Mn site leading to reduction of the effective magnetic moment. Simultaneously transitions between continuum band states and quasibound states define spin fluctuations of such composite magnetic moment\cite{Demishev11,Demishev12}. This quasibound state may be viewed as an analogue of spin polaron, and it is possible to show that the model based on these specific states allows explaining peculiarities of magnetic scattering and magnetic resonance data better than traditional itinerant approach\cite{Demishev11,Demishev12}. 

Therefore it is interesting to consider the results on the magnetization structure obtained in the present study in the framework of spin polaron model\cite{Demishev11,Demishev12}. At first glance in the Heisenberg paradigm without any spin polarons, there are LMMs on Mn sites and free electrons. Therefore it is possible to speculate that saturating part of magnetization is due to LMMs subsystem, whereas linear part is nothing but enhanced Pauli term. However, the estimates of contributions to magnetization from LMMs and free electrons show that magnitude of the possible Pauli term is about three orders of magnitude less than that of LMMs\cite{Demishev12}. Consequently the aforementioned explanation must imply an increase of the effective mass by a factor $\sim 10^2-10^3$, which does not meet the experimental situation in MnSi\cite{Mena03}.

In our opinion, this problem may be resolved when possible composite nature of LMM in Mn$_{1-x}$Fe$_x$Si, which follows from spin polaron model, is taken into account. If only renormalized (screened by quasibound electrons) manganese LMMs are considered, and they do not depend on the magnetic field, there should be one saturating term in magnetization. Since the infinite magnetic field always lead to a parallel alignment of all spins in the sample and in zero magnetic field spin polaron state is characterized by antiparallel orientation of the LMM and spin of quasibound electron, the increase of magnetic field may induce gradual changes of spin alignment resulting in additional increase of magnetization. In the model considered in Refs.~\onlinecite{Demishev11,Demishev12} the corresponding process has an on-site character and may give rise to additional linear term in magnetization. Assuming certain hierarchy of magnetic interactions, when lowest in Zeeman energy is the process of spin polarization of considered composite LMM as a whole and the change of spins orientation inside spin polarons is characterized by strongest interaction energy, it is possible to come to magnetization structure expressed by Eq.~(1) at least as a reasonable approximation.

The interpretation regarded in Refs.~\onlinecite{Demishev11,Demishev12} allows explaining reduced value of $M_0$. The temperature $T_s$ may be treated as an analogue of Curie temperature for the phase with short-range magnetic order\cite{Grigoriev11} and thus this parameter reflects characteristics of interaction \emph{between} spin polarons in the mean field approximation. However, a difference between $T_s(x)$ and $T^*(x)$ (for $x>x^*$ these parameters have even opposite signs) suggest the presence of another type of magnetic interaction different from that describing ordering of spin polarons. Moreover, the Curie-Weiss form for the additional term in Eq.~(8) may suggest that this additional interaction can be also described by some mean field. Apparently it is not the case for the simple on-site model\cite{Demishev11,Demishev12}, however it may become possible, when the existence of a small cluster, where the electron spins are oriented opposite to the localized magnetic moments of Mn and all manganese LMMs are aligned in the same direction due to strong ferromagnetic coupling, is assumed. 

If LMMs and electrons may be treated in quasiclassical approximation this cluster is a kind of \textquotedblleft elementary\textquotedblright ferrimagnet. It is worth noting that in some ferrimagnets like Mn[FeCr]O$_4$ the magnetization field dependence  contains both saturating and non-saturating linear parts\cite{Vonsovskii74}. In this case the characteristic temperature $T^*(x)$ corresponds to the mean field inside the ferrimagnetic cluster and the iron doping induced change of the interaction parameters between band electrons and manganese LMMs may alter and even change the sign of the effective paramagnetic temperature in the corresponding Curie-Weiss law as in conventional ferrimagnets\cite{Vonsovskii74}. At the same time, $T_s(x)$ will characterize interaction between the ferromagnetic spin clusters.

The problem which poses serious difficulties for considered ansatz is the enhanced value of $\mu^*\sim 6~\mu_B$. Indeed, LDA calculations give the magnitude of the bare Mn magnetic moment $\mu_{Mn}\sim 1.2~\mu_B$\cite{Corti07}. In order to explain this discrepancy, we suppose that the manganese LMMs inside a cluster are fixed in parallel alignment , which is equivalent of the enhanced magnetic dipole formation with the magnitude $\sim N_{Mn}\mu_{Mn}$ (here $N_{Mn}$ denotes number of Mn ions belonging to the spin cluster). This dipole is screened by itinerant electrons inside cluster and corresponding saturated magnetization can be estimated as $M_0\sim N_{Mn}\mu_{Mn}-n_e\mu_e$, where $n_e$ and $\mu_e$ are average number of electrons in cluster and their effective magnetic moment respectively. As indicated above, in Mn$_{1-x}$Fe$_x$Si screening of coupled Mn LMMs occurs via quasibound electron states. Additionally there are frequent transitions between these states and continuum band states serving as a source of spin fluctuations\cite{Demishev11,Demishev12}. In the case of strong spin fluctuations, it is possible to expect that the orientations of the electron spins will just follow the Mn LMMs subsystem spin polarization, which dependence on the magnetic field is controlled by enhanced magnetic dipole. Neglecting possible renormalization of the magnetic dipole magnitude by the spin fluctuations it is capable of estimate the number of Mn LMMs in a cluster as $N\sim\mu^*/\mu_{Mn}\sim 5$. Consequently the \textquotedblleft elementary\textquotedblright ferrimagnet is expected to have a size of about the unit cell and can be considered as a ferrimagnetic nanodroplet (there are from two to three Mn ions in the unit cell of Mn$_{1-x}$Fe$_x$Si for $x<0.3$ (see the Ref.~\onlinecite{Grigoriev10})). 

The considered interpretation of Eq.~(8) and the corresponding parameters looks very unusual and assumes magnetic inhomogeneity of the paramagnetic phase on the nanoscale. Interesting that qualitatively same supposition was applied recently for an explanation of the $\mu$SR experiments in MnSi\cite{Storchak11}. Nevertheless the realness of the proposed mechanism requires development of a quantitative theory, which may clarify physical ground leading to universal scaling of magnetization in the paramagnetic phase in Mn$_{1-x}$Fe$_x$Si solid solutions and the exact meaning of the empirical parameters with the analytic approximation. In this connection, it is possible to mark recent work\cite{Dmitrienko12}, where appearance  of antiferromagnetic correlations (negative sign of $T^*$ in our experiments) induced by Dzyaloshinskii-Moriya interaction in MnSi-type solids was predicted.

\section{Conclusions}

In conclusion, we showed that in the range $B<5$~T and $T<60$~K field dependences of the magnetization $M(B,T=const)$ in the paramagnetic phase of Mn$_{1-x}$Fe$_x$Si solid solutions $(x<0.3)$ exhibit scaling behaviour of the form $B^2\partial M/\partial B-M=F(B/(T-T_s))$, where $T_s$ denotes empirically determined temperature of the transition into magnetic phase with fluctuation driven short-range magnetic order and $F(\xi)$ is a universal scaling function for given composition. The scaling relation allowed concluding that magnetization in the paramagnetic phase of Mn$_{1-x}$Fe$_x$Si is represented by the sum of two terms. The first term of the sum is saturated by the scaling variable $\xi=B/(T-T_s)$, whereas the second is  linearly dependent on the magnetic field. A simple analytical formula describing magnetization is derived (Eq.~(8)) and applied to estimates of the parameters characterizing localized magnetic moments in the studied system. The obtained data may be qualitatively interpreted assuming magnetic inhomogeneity of the paramagnetic phase on the nanoscale.

\textit{\textbf{Acknowledgements.}} This work was supported by Programme of Russian Academy of Sciences \textquotedblleft Strongly correlated electrons\textquotedblright and by RFBR grant 13-02-00160. Authors are grateful to S.M.~Stishov for helpful discussions.

\end{document}